\documentclass[12pt,final]{preprint}

\usepackage{cite}
\usepackage{epsfig}
\usepackage{amssymb} % use 'amssymb' before our 'symbols'
\usepackage{symbols}

\newcommand{\vers}{15.05.95}

\newcommand{\chup}{$\chi U\phi$}

\newcommand{\chupiv}{$\chi U\phi_4$}
\newcommand{\cbcex}{\langle\overline{\chi}\chi \rangle}

\newcommand{\scsb}{S$\chi$SB}
\newcommand{\pdp}{\phi^{\dagger}\phi}
\newcommand{\pdc}{\phi^{\dagger}\chi}

\begin{document}
%%%%%%%%%%%%%%%%%%%%%%%%%%%%%%%%%%%%%%%%%%%%%%%%%%%%%%%%%%%%%%%%%
\dateandnumber(May 1995){J{\"u}lich, HLRZ 23/95}%
%%%%%%%%%%%%%%%%%%%%%%%%%%%%%%%%%%%%%%%%%%%%%%%%%%%%%%%%%%%%%%%%%
\titleofpreprint%
{                   Chiral phase transition in                 }%
{              a lattice fermion--gauge--scalar model          }%
{              with U(1) gauge symmetry$^*$                    }%
{                                                              }%
{                                                              }%
%%%%%%%%%%%%%%%%%%%%%%%%%%%%%%%%%%%%%%%%%%%%%%%%%%%%%%%%%%%%%%%%%
\listofauthors%
{  W.~Franzki$^{1,2,3}$, C.~Frick$^{1,4}$,    }%
{    J.~Jers{\'a}k$^{1,2,5}$ and X.Q. Luo$^{2,6}$                                            }%
{                                                              }%
%%%%%%%%%%%%%%%%%%%%%%%%%%%%%%%%%%%%%%%%%%%%%%%%%%%%%%%%%%%%%%%%%
\listofaddresses%
{\em $^1$Institute of Theoretical Physics E,
  RWTH Aachen,         D-52056 Aachen, Germany              }%
{\em $^2$HLRZ c/o KFA J{\"u}lich,
      D-52425 J{\"u}lich, Germany                                }%
{
                                                               }%
{}
%%%%%%%%%%%%%%%%%%%%%%%%%%%%%%%%%%%%%%%%%%%%%%%%%%%%%%%%%%%%%%%%%

\abstractofpreprint{The chiral phase transition induced by a charged scalar
  field is investigated numerically in a lattice fermion-gauge-scalar model
  with U(1) gauge symmetry, proposed recently as a model for dynamical fermion
  mass generation.  For very strong gauge coupling the transition is of second
  order and its scaling properties are very similar to those of the
  Nambu--Jona-Lasinio model.  However, in the vicinity of the tricritical
  point at somewhat weaker coupling, where the transition changes the order,
  the scaling behavior is different.  Therefore it is worthwhile to
  investigate the continuum limit of the model at this point.}

%%%%%%%%%%%%%%%%%%%%%%%%%%%%%%%%%%%%%%%%%%%%%%%%%%%%%%%%%%%%%%%%%
\footnoteoftitle{
%%%%%%%%%%%%%%%%%%%%%%%%%
\footnoteitem($^*$){ \sloppy
Supported
by Deutsches Bundesministerium f{\"u}r Forschung und Technologie and
by Deu\-tsche Forschungsgemeinschaft.
}
%%%%%%%%%%%%%%%%%%%%%%%%%
\footnoteitem($^3$){ \sloppy
E-mail address: w.franzki@kfa-juelich.de
}
\footnoteitem($^4$){ \sloppy
Present address: AMS, Querstr. 8-10, D-60322 Frankfurt a.M., Germany
}
\footnoteitem($^5$){ \sloppy
E-mail address: jersak@physik.rwth-aachen.de
}
\footnoteitem($^6$){ \sloppy
E-mail address: luo@hlrz.kfa-juelich.de
}%%%%%%%%%%%%%%%%%%%%%%%%%
}
%%%%%%%%%%%%%%%%%%%%%%%%%%%%%%%%%%%%%%%%%%%%%%%%%%%%%%%%%%%%%%%%%
%
\pagebreak
%\tableofcontents
%
%1111111111111111111111111111111111111111111111111111111111111111111111
%%%%%%%%%%%%%%%%%%%%%%%%%%%%%%%%%%%%%%%%%%%%%%%%%%%%%%%%%%%%%%%%%%%%%%%
%%%%%%%%%%%%%%%%%%%%%%%%%%%%%%%%%%%%%%%%%%%%%%%%%%%%%%%%%%%%%%%%%%%%%%%
\section{Introduction}
Scalar fields are mostly introduced into the quantum field theoretical models
in order to trigger some symmetry breaking already on the tree level, as in
the conventional Higgs mechanism.  However, fundamental scalar fields can also
play an important role in a dynamical symmetry breaking occurring only beyond
the perturbative expansion.  An example are strongly coupled Yukawa models,
which can exhibit spontaneous chiral symmetry breaking (\scsb) even if the
bare scalar potential in the action does not have the form of the classical
Mexican hat~\cite{KoTa90,Ca91,ClRo91,BoDe92b}.

In this paper we continue the investigation of another model of this type, the
lattice chiral symmetric \chupiv\ model, considered in this context in
ref.~\cite{FrJe95a}.  The model consists of staggered fermion field $\chi$,
strongly coupled gauge field~$U$ and scalar field~$\phi$ in four dimensions
(4D).  The gauge symmetry is compact U(1), $U\in {\rm U(1)}$, and the model is
confining at strong coupling.  Both matter fields have unit charge and are,
consequently, confined.  In one phase, the \emph{Nambu phase}, \scsb\ is
generated dynamically by the vectorlike interaction between the gauge
field~$U$ and the fermion field~$\chi$, making the gauge invariant
condensate~$\cbcex$ nonzero, in analogy to the \scsb\ in QCD. The Goldstone
bosons are composed of $\chb$ and $\chi$.

The role of the fundamental charged scalar field~$\phi$ is twofold.  First, it
shields the U(1)-charge of the fundamental fermion $\chi$, so that a composite
neutral physical fermion state of the form~$F = \pdc$ can exist asymptotically
in spite of the confinement of the U(1)-charge.  In the Nambu phase its
mass~$m_F$ is generated dynamically.  This mechanism of fermion mass
generation has been called {\em shielded gauge mechanism} in ref.
\cite{FrJe95a}.

Second, the scalar field has the tendency to suppress the \scsb\ and
for very strong gauge coupling induces a new second order phase transition,
at which the chiral  symmetry is restored and $am_F$, the fermion
mass in lattice units, continuously approaches zero. 
In the scaling region of this phase transition in the Nambu phase,
the model would describe a massive fermion in the continuum
if it was nonperturbatively renormalizable.

We initiate the investigation of this crucial question by means of numerical
simulations of the \chupiv\ model with dynamical fermions. The chiral phase
transition is studied by varying suitably the gauge coupling $\beta=1/g^2$,
the hopping parameter $\kappa$ and the bare fermion mass $am_0$. Previous
studies of the present
\cite{LeShr87a,LeShr87b,LeShr88b,Shr89,DaKo88,FrJe95a,FrJe95b,LuFr95,FrLu95}
and similar \cite{LeShi86b,DeShi88,AoLe88,Ku89a,DaMe89,MePe91} models have
indicated that the chiral phase transition, at which $am_F$ vanishes, is
smooth in some interval of strong gauge coupling, including $\beta = 0$.  At
some value $\beta_{\rm E}$ of $\beta$ this line changes the order at the
\emph{tricritical point} E and continues as a first order transition line for
larger $\beta$.  We confirm these earlier results in high statistics simulations
and localize with good accuracy the phase transition line and less precisely
the point E ($\beta_{\rm E} \simeq 0.64$) on it.  We investigate the scaling
behavior of several observables in the vicinity of the transition.

At $\beta=0$, the \chupiv\ model is identical to the 4D Nambu--Jona-Lasinio
(NJL) model on the lattice with the same global chiral symmetry.  That model
has been investigated recently in much detail \cite{AlGo95} and its
nonrenormalizability has been confirmed. We compare data obtained for $\beta >
0$ on lattices of various sizes and at different values of the bare fermion
mass $am_0$ with those from the NJL model, using analogous analysis methods.
We find that in the interval $0 < \beta < \beta_{\rm E} - \varepsilon$,
$\varepsilon \lsim 0.1$, the behavior of the \chupiv\ model is very similar to
that of the NJL model.

In the close vicinity of the tricritical point E the behavior changes
significantly, however.  The analysis methods, that work well at smaller
$\beta$-values, fail to describe the finite size effects and the
$am_0$-dependence of the data.  Furthermore, a new correlation length, which
is of the order of the lattice spacing or less at smaller $\beta$-values, gets
large here, indicating the presence of a new state in the spectrum of the
model. This scalar bound state $S = \pdp$ has no counterpart in the NJL model.

From these observations we tentatively conclude that there are two
possibilities how to approach the continuum limit of the \chupiv\ model from
within the Nambu phase in such a way that the fermion mass $am_F$ scales, and
a massive fermion $F$ is thus found in the continuum limit:
\begin{enumerate}
\item At some point on the chiral phase transition line at $\beta < \beta_{\rm E}$
  (perhaps even at negative $\beta$).  Here some generalization of the NJL
  model is obtained, possibly similar to those discussed in
  ref.~\cite{HaHa91}.  Either it is a nonrenormalizable theory like the NJL
  model itself, or a Yukawa-like model.
\item At the tricritical point E.  The spectrum would contain a massive
  fermion $F$, a massive scalar $S$ and the obligatory Goldstone bosons $\pi$.
  It could well be that the model is renormalizable and that the shielded
  gauge mechanism works here, as conjectured in ref.~\cite{FrJe95a}. Then the
  question would be, whether the resulting theory is Yukawa-like or whether
  some new interesting universality class exists here.
\end{enumerate}
Until now we have been able to localize the tricritical point E and to analyze
the properties of the chiral phase transition outside the scaling region of
this point.  We found that the same methods of analysis do not work in the
vicinity of E.  The understanding of the point E itself requires substantially
more data and new analytic tools for their analysis and is beyond the scope of
this work.

The outline of the paper is as follows: In the next section we define the
lattice \chupiv\ model and describe its phase structure at strong gauge
coupling.  Here we collect the quantitative results on the position of the
phase transition lines (see fig.~\ref{PD}).  In sects.~3 and 4 we then
describe in some detail the observed properties of the first order part ($
\beta_{\rm E}<\beta<\beta_{\rm T} \simeq 0.9$) and second order part ($\beta <
\beta_{\rm E}$), respectively, of the chiral phase transition line.  Our
current preliminary experience with the vicinity of the point E is presented
in sect.~5.  Sect.~6 contains summary and outlook.  In the appendix we
describe another tool used in this paper in addition to the hybrid Monte Carlo
method: the microcanonical fermionic average method developed in
refs.~\cite{AzDi90,AzLa93} and extended to the \chupiv\ model \cite{Lu95}.

%2222222222222222222222222222222222222222222222222222222222222222222222
%%%%%%%%%%%%%%%%%%%%%%%%%%%%%%%%%%%%%%%%%%%%%%%%%%%%%%%%%%%%%%%%%%%%%%%
%%%%%%%%%%%%%%%%%%%%%%%%%%%%%%%%%%%%%%%%%%%%%%%%%%%%%%%%%%%%%%%%%%%%%%%
\section{Lattice \chupiv~model at strong coupling}

\subsection{The \chupiv~model}
%%%%%%%%%%%%%%%%%%%%%%%%%%%%%%%%%%%%%%%%%%%%%%%%%%%%%%
%
The \chupiv\ model is defined on the euclidean hypercubic lattice in 4D as
follows: The staggered fermion field $\chi$ of charge one leads to the global
U(1) chiral symmetry of the model in the chiral limit, i.e. when the bare
fermion mass $m_0$ vanishes.  The gauge field link variables $U$ are elements
of the compact gauge group U(1).  The complex scalar field $\phi$ of charge one
satisfies the constraint~$|\phi |$=1.

The action is
\begin{equation}
    S_{\chi U \phi} = S_\chi + S_U + S_\phi \; ,
\lb{action}
\end{equation}
where
\begin{eqnarray}
  S_\chi &=& {\textstyle \half} \sum_x \sum_{\mu = 1}^4
             \eta_{\mu x} \chb_x \left[ U_{x,\mu} \chi_{x + \mu} -
                     U_{x-\mu,\mu}^\dagger  \chi_{x - \mu} \right]
              \, + \, a m_0 \sum_x \chb_x \chi_x \; ,
\lb{SCH}                  \\
  S_U   &=& \beta \, \sum_{\rm P} \left[ 1 - \mbox{Re} \{ U_{\rm P} \} \right] \; ,
\lb{SU}  \\
  S_\phi &=& - \kp \, \sum_x \sum_{\mu=1}^4 \left[ \phi_x^\dagger
              U_{x,\mu} \phi_{x + \mu} \,+\, {\rm h.c.} \right] \; .
\lb{SPH}
\end{eqnarray}
Here $U_{\rm P}$ is the plaquette product of link variables
$U_{x,\mu}$ and $\eta_{\mu x} = (-1)^{x_1 + \cdots + x_{\mu - 1}}$.
The hopping parameter~$\kp$ vanishes (is infinite) when the squared bare
scalar mass is positive (negative) infinite.
The bare fermion~mass~$m_0$ is introduced for technical reasons, and the
model is meant in the limit~$m_0 \!=\! 0$.

\subsection{Equivalence of the $\chi U \phi$ model to the
  four fermion theory in the strong coupling limit}
%%%%%%%%%%%%%%%%%%%%%%%%%%%%%%%%%%%%%%%%%%%%%%%%%%%%%%
%
At~$\beta \!=\! 0$, the \chup~model in~$d$ dimensions can be rewritten exactly
as a lattice four fermion model~\cite{LeShr87a}.  In the path integral of the
model~(\ref{action}) the scalar and gauge fields can be integrated out.  This
results in a purely fermionic model with the action
\begin{equation}
   S_{\rm 4f} = - \sum_x \sum_{\mu = 1}^d \left[ \, G \;
              \chb_x \chi_x \chb_{x+\mu} \chi_{x+\mu} \, - \,
             {\textstyle \half} \eta_{\mu x}
             \left( \chb_x \chi_{x+\mu} - \chb_{x+\mu} \chi_x \right)
             \right]  \, + \, \frac{am_0}{r} \sum_x \chb_x \chi_x \; .
\lb{action4f}
\end{equation}
Here
\begin{equation}
\renewcommand{\arraystretch}{1.4}
   r = r(\kp) = \frac{J_U}{J_1}  \;\;\;\;\;\; \mbox{ and } \;\;\;
       \begin{array}{rclll}
         J_U &=& \int {\rm d} U e^{2 \kp \Reop \{ U \} } U &=&
          I_1(2\kp) \;, \\
         J_1 &=& \int {\rm d} U e^{2 \kp \Reop \{ U \} }   &=&
          I_0(2\kp) \;.
       \end{array}
\renewcommand{\arraystretch}{1.}
\lb{rofkappa}
\end{equation}
The fermion field has been rescaled by~$\sqrt{r}$.
The parameter $r$ is an analytic function of~$\kp$ increasing monotonically
from~$r(0) \!=\! 0$ to~$r(\infty) \!=\! 1$.

The action (\eq{action4f}) obviously describes a lattice version of the
NJL model.
The four~fermion coupling parameter~$G$ is related to~$\kp$ via~$r$:
\begin{equation}
   G = \frac{1-r^2}{4 r^2} \; .
\lb{GNJL}
\end{equation}
From (\eq{rofkappa}) one sees that~$G$ is decreasing monotonically
with increasing~$\kp$;~$G \!=\! \infty$ at~$\kp \!=\! 0$,
and~$G \!=\! 0$ at~$\kp \!=\! \infty$.

%========================================================
\subsection{Phase diagram of the \chupiv~model}
%%%%%%%%%%%%%%%%%%%%%%%%%%%%%%%%%%%%%%%%%%%%%%%%%%%%%
%
%FFFFFFFFFFFFFFFFFFFFFFFFFFFFFFFFFFFFFFFFFFFFFFFFFFFFFFFFFFFFFFFFFFFFF
%%%%%%%%% phase diagram of chi-U-phi_4 model at strong coupling %%%%%%
%%%%%%%%%%%%%%%%%%%%%%%%%%%%%%%%%%%%%%%%%%%%%%%%%%%%%%%%%%%%%%%%
\begin{figure}%[t]
  \centerline{\hspace{8mm}
    \epsfig{file=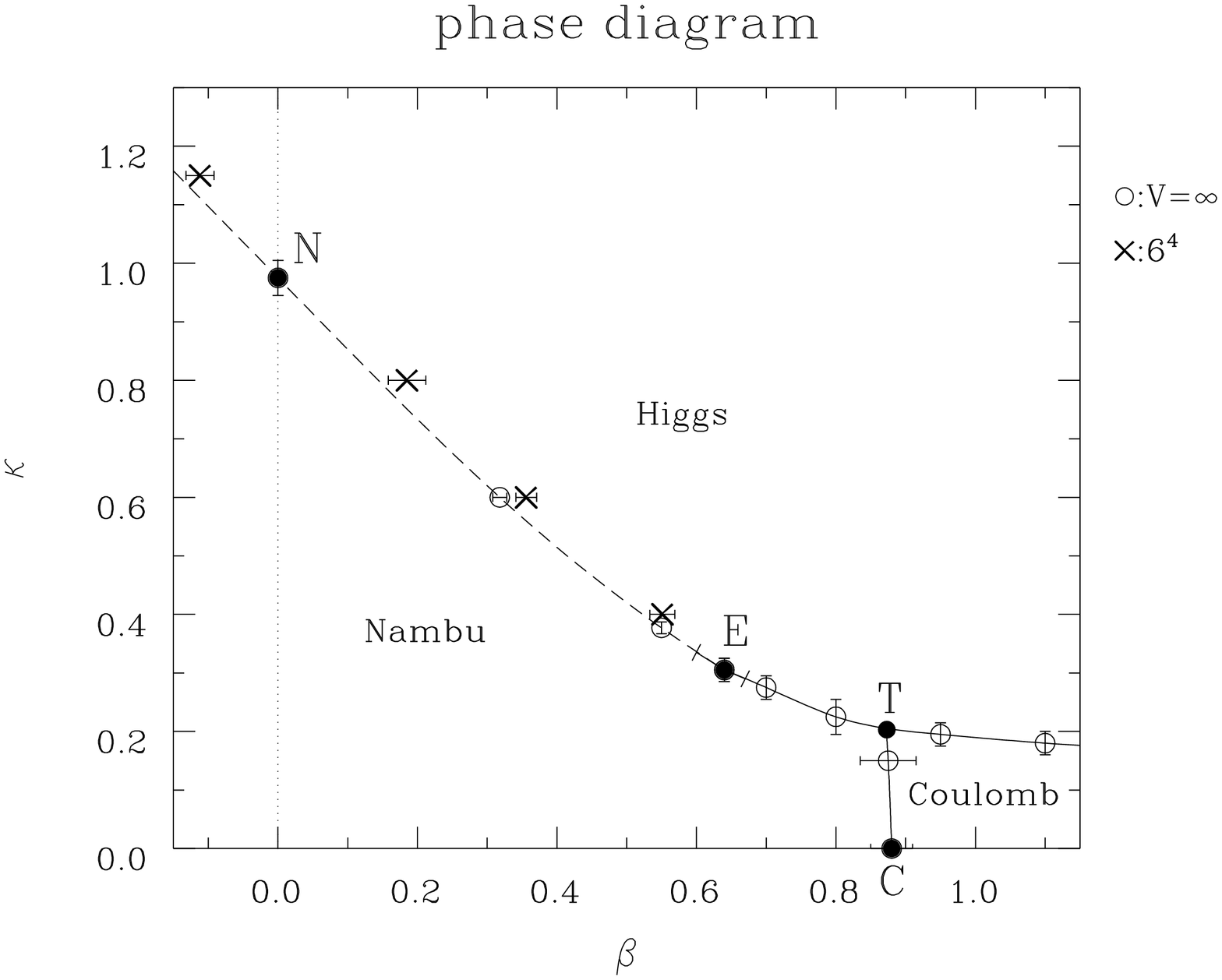,width=13cm,angle=90}}
  \vspace{-10mm}%
\caption[see below the figure]{%
  The phase diagram of the~\chupiv~model~defined in
  eq.~({\protect\ref{action}}) at $m_0 = 0$.  Three phases, the Coulomb, Higgs
  and Nambu phases, are found for $\kappa \geq 0$ and $\beta \geq 0$ or $\beta
  \lsim 0$.  The emphasized points are:\\*[3mm]
  \begin{minipage}{\textwidth}
    \begin{list}{}{\itemsep0pt \parsep0pt}
    \item[\bf N:] critical point of the {\bf N}JL~model, which is a special
      case of the~\chupiv\ model at~$\beta \!=\! 0$,
    \item[\bf E:] critical {\bf E}ndpoint of the Higgs phase transition line,
    \item[\bf T:] {\bf T}riple point,
    \item[\bf C:] phase transition from the {\bf C}onfinement (at strong gauge
      coupling) to\\ the {\bf C}oulomb phase (at weak gauge coupling) in the
      model without the scalar field.
    \end{list}
  \end{minipage}
  \\*[3mm]%
  The dashed line corresponds to a 2$^{\rm nd}$ order phase~transition, full
  lines to 1$^{\rm st}$ order transitions.  The dynamical fermion mass
  generation takes place in the Nambu phase.  A detailed discussion of this
  phase diagram is given in ref.~\protect\cite{FrJe95a}. \\*[3mm]%
  The circles represent extrapolations of pseudocritical points to infinite
  volume and chiral limit.  Also shown are the results obtained by means of
  the microcanonical fermionic average method on a $6^4$-lattice (crosses) in
  the chiral limit.}
\label{PD}
\end{figure}
%%%%%%%%%%%%%%%%%%%%%%%%%%%%%%%%%%%%%%%%%%%%%%%%%%%%%%%%%%%%%%%%%%%%%%%%%%%%%%
%%FFFFFFFFFFFFFFFFFFFFFFFFFFFFFFFFFFFFFFFFFFFFFFFFFFFFFFFFFFFFFFFFFFFFFFFFFFFF
\begin{table}
  \begin{center}
    \leavevmode
    \begin{tabular}[c]{|l|l|}
      \hline
      \hspace{5mm}$\beta$ & \hspace{5mm}$\kappa$ \\ \hline \hline
      0.00 & 0.975(30) \\
      0.318(10) & 0.60 \\
      0.55 & 0.377(10)\\
      0.64 & 0.305(20)\\
      0.70 & 0.275(20)\\
      0.80 & 0.225(30)\\ \hline
      0.95 & 0.195(20)\\
      1.10 & 0.180(20)\\ \hline
      0.875(40) & 0.15 \\
      0.880(30) & 0.00 \\\hline
    \end{tabular}
    \caption{Estimated position of the phase transitions in the chiral limit
      and on infinite volume.}
    \label{PDtab}
  \end{center}
\end{table}

The phase diagram of the \chupiv~model with U(1) gauge symmetry obtained in
this work is shown in fig.~\ref{PD}, and the positions of the phase
transitions are listed in table~\ref{PDtab}. The used methods are described in
the next sections. The phase diagram is consistent with the earlier results of
unquenched simulations~\cite{DaKo88,LuFr95,FrLu95}. Qualitatively it agrees
also with the quenched simulations~\cite{LeShr87b,LeShr88b,Shr89,FrJe95a}. The
major difference is that the points E, T and C lie at smaller $\beta$ values
than in the quenched case.

The chiral phase transition of the NJL~model at~$\bt\!=\!0$ (point~N), seen in
the chiral condensate and the fermion mass $am_F$ extends to nonzero values
of~$\bt$.  As pointed out in ref.~\cite{LeShr87b}, this can be derived by
means of a convergent expansion around $\bt = 0$.  Therefore, some properties
of the NJL model have been expected to persist also at small nonzero $\bt$.
We find this is so until $\bt \simeq 0.55$.

The model exhibits the Higgs phase transition (line ET and its nearly
horizontal continuation to larger $\beta$) seen in bosonic observables like
plaquette and link energies.  The striking feature is that, within the
numerical precision, the chiral phase transition joins the Higgs phase
transition line~ET at the tricritical point~E, forming a smooth line~NET.  The
NE and ET parts are of second and first order, respectively.  To our knowledge
the interweaving of the chiral and Higgs phase transition is not understood
theoretically.

We concentrate on the Nambu phase, which is the area below the NET~line.
Here the chiral condensate~$\cbcex$ is nonvanishing.
Both~$\chi$ and~$\phi$ fields are confined, in analogy to the quark
confinement in  QCD.
The mass~$am_F$ of the fermion state~$F = \pdc$ is nonzero,
and thus the dynamical mass generation occurs.

The other phases as well as various limit cases of the phase diagram are
discussed in much detail in ref.~\cite{FrJe95a}.  It is helpful to keep in
mind that in the quenched approximation the phase diagram corresponds to that
of the U(1) Higgs model (the present model without fermion field), which has
been thoroughly investigated \cite{JaJe86,EvJa87b,CaPe87,AlAz9293}. Even in
this approximation, the chiral phase transition is observable in the chiral
condensate, which is a nonlocal function of bosonic fields.

\subsection{Definitions of observables}
%%%%%%%%%%%%%%%%%%%%%%%%%%%%%%%%%%%%%%%%%%%%%%%%%%%%%%%%%%%%%%%%%%%%%%%%
%
For a localization of the phase transition lines and a determination of the
particle spectrum the following observables are used:

The normalized plaquette and link energies, defined as
\begin{eqnarray}
    E_{\rm P} &=& \frac{1}{6V} \sum_{\rm P}
                    \Reop \{ U_{\rm P} \} \;, \lb{EP} \\
    E_{\rm L} &=& \frac{1}{4V} \sum_{x,\mu}
                    \Reop \{ \phi^\dagger_x U_{x,\mu} \phi_{x+\mu} \} \;,
\lb{EL}
\end{eqnarray}
where~$V\!=\!L^3 T$ is the lattice volume.
These observables are useful for the localization of
the Higgs phase transition.
It is convenient \cite{AlAz9293} to use the perpendicular component of these energies,
$E_\perp$ defined as
\begin{equation}
E_\perp = E_{\rm L} \cos{\theta} + E_{\rm P} \sin{\theta},
\lb{E-PERP}
\end{equation}
$\theta(\bt)$ being the slope of the chiral phase transition line at point
$\bt$.

For the localization of the chiral phase transition line
we measure the chiral~condensate $\cbcex$ with the stochastic~estimator~method.

Further we introduce the quantity
\begin{equation}
  \Omega = \frac{2}{V} \left.\left\langle \sum_{i=1}^{V/2} {1 \over
          \lambda_i^2} \right\rangle\right\vert_{m_0=0} \;,
\lb{eq:omega}
\end{equation}
where $\lambda_i$ are the positive eigenvalues of the massless fermion matrix.
This quantity equals the chiral susceptibility in the chiral symmetric phase.
Of course, in the phase with \scsb\ it differs from the susceptibility,
actually it diverges as $O(V)$. However, $\Omega$ can be rather easily
calculated on small lattices by means of the microcanonical fermionic average
method, explained in the appendix, and used for an estimate of the position of
the pseudocritical point at $m_0=0$.

To calculate the mass $am_F$ of the physical fermion we consider the gauge
invariant fermionic field
\begin{equation}
   F_x = \phi^\dagger_x \chi_x \;, \hspace*{2cm}
   \overline{F}_x = \phi_x \chb_x \; ,
\lb{FFbar}
\end{equation}
and determine numerically the corresponding fermion propagator in momentum
space
\begin{equation}
  G_{F\,AB}(p) = \left\langle \frac{1}{V} \sum_{x,y} e^{i(p+\pi_A)x}
    F_x \overline{F}_y e^{-i(p+\pi_B)y} \right\rangle\,.
\end{equation}
The $\pi_A$, as well as the $\Gamma_\mu$ below, are defined in
refs.~\cite{GoSm84,DoSm83}.  We look for the projections
\begin{equation}
  \left. G^\Gamma_\mu(p_t) = i {\textstyle \frac{1}{16}} \tr
  \left\{G_F(\vec{0},p_t)\Gamma_\mu\right\}\right|_{\mu=t} \;, \hspace{7mm}
  G^\un(p_t) = {\textstyle \frac{1}{16}} \tr G_F(\vec{0},p_t) \;,
\end{equation}
which we fit by the formulas
\begin{eqnarray}
  G^\Gamma_t(p_t) &=& \frac{Z_F\sin p_t}{\sin^2 p_t+(am_F)^2}
  \lb{fitmf1} \\
  G^\un(p_t) &=& \frac{Z_F am_F}{\sin^2 p_t+(am_F)^2} \,.
  \lb{fitmf2} 
\end{eqnarray}
Both fits and also those in configuration space give consistent results. The
presented results are determined by means of eq.~(\ref{fitmf1}).

Further we consider the fermion-antifermion composite states, the ``mesons''.
The time\-slice operators for the mesons are given by
\begin{equation}
   {\cal O}^{ik} (t) = \sum_{\vec{x}} s^{ik}_{\vec{x},t} \chb_{\vec{x},t}
                        \chi_{\vec{x},t} \;,
\lb{Omesons}
\end{equation}
with the sign factors~$s^{ik}_x$ and other details given in
ref.~\cite{FrJe95a}.  We measure their correlation functions using point
sources.

We are also interested in the scalar and the vector bosons, which are present also
in the gauge-scalar model without fermions.  The corresponding
operators~\cite{EvJa87b} are
\begin{eqnarray}
    {\cal O}^{(S)} (t) &=& \frac{1}{L^3} \sum_{\vec{x}} \Reop
              \left\{
       \sum_{i=1}^{3} \phi^\dagger_{\vec{x},t} U_{(\vec{x},t), i}
                      \phi_{\vec{x}+\vec{i},t} \right\} \;,
\lb{OS} \\
    {\cal O}^{(V)}_{i} (t) &=& \frac{1}{L^3} \sum_{\vec{x}} \Imop
              \left\{ \phi^\dagger_{\vec{x},t} U_{(\vec{x},t), i}
                      \phi_{\vec{x}+\vec{i},t} \right\} \;, \;\;\;\;\;\;
                             i = 1,\;2,\;3 \;.
\lb{OV}
\end{eqnarray}
The masses of the scalar boson~$a m_S$ and the vector boson~$a m_V$ are
computed from the corresponding correlation functions in momentum space. We
note that the states $S$ and $V$ are expected to mix with the mesons carrying
the same quantum numbers. In the QCD language these would be the $\sigma$ and
$\om$ mesons, respectively.

%33333333333333333333333333333333333333333333333333333333333333333333333
%%%%%%%%%%%%%%%%%%%%%%%%%%%%%%%%%%%%%%%%%%%%%%%%%%%%%%%%%%%%%%%%%%%%%%%%
%%%%%%%%%%%%%%%%%%%%%%%%%%%%%%%%%%%%%%%%%%%%%%%%%%%%%%%%%%%%%%%%%%%%%%%%
\section{ET line}
The ET line to the right of the point E (fig.~\ref{PD}) is a first order phase
transition line, and a continuum limit cannot be constructed here.  We have
investigated it mainly in order to verify the coincidence of the chiral and
Higgs phase transitions.  A better understanding of the ET line is also
helpful for the localization and investigation of the tricritical point E.

As known from the studies of the U(1) Higgs model \cite{JaJe86,AlAz9293}, the
ET line (Higgs transition) is in the quenched approximation of first order in
bosonic variables $E_\perp, am_S$ and $am_V$.  In this approximation also the
fermionic observables $\cbcex$ and $am_F$ show discontinuities at the same
values of the coupling parameters as the bosonic variables do.  We have found
similar discontinuities in the full model (i.e.\ with dynamical fermions) too.
\begin{figure}%[t]
  \centerline{\epsfig{file=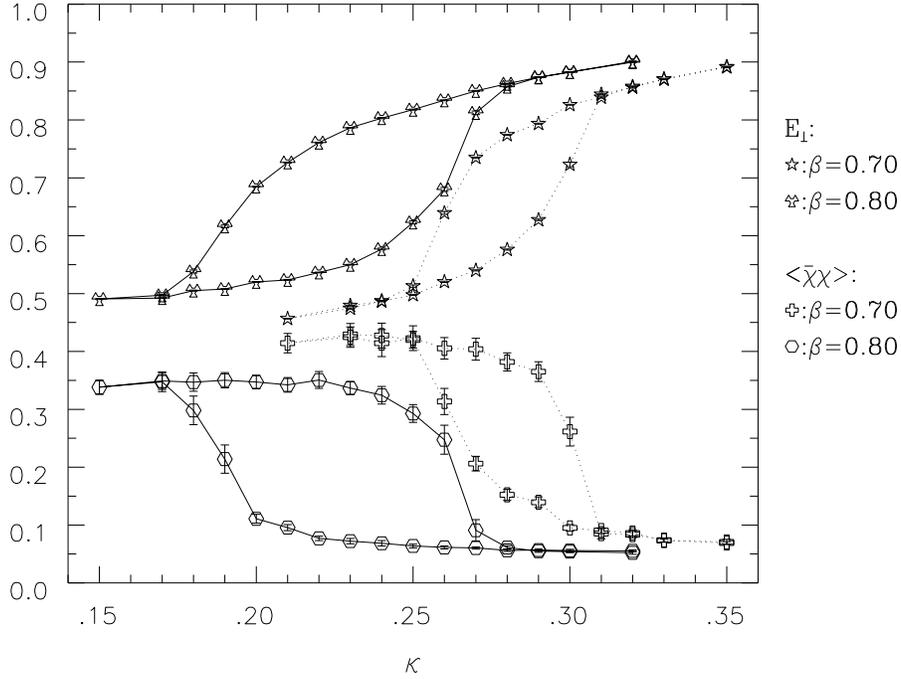,width=11cm,angle=90}}
  \vspace{-10mm}%
  \caption{%
    Thermal cycles at $\beta=0.70$ and $\beta=0.80$. Both $E_\perp$ and
    $\cbcex$ have hystereses at the same place.}
  \label{fig:HIST}
\end{figure}
This is demonstrated for $\beta=0.70$ and $\beta=0.80$ on the $8^4$ lattice for
$am_0=0.04$ in fig.~\ref{fig:HIST}. Here $E_\perp$ and $\cbcex$ develop
hystereses in thermal cycles in the $\kappa$ direction with low statistics per
point (100 trajectories).

The positions of the hystereses in bosonic and fermionic observables,
$E_\perp$ and $\cbcex$ respectively, apparently coincide. This is so also for
other values of $am_0$ and on a $6^4$~lattice. The hystereses shift slightly
to lower $\kappa$-values as $am_0$ is decreased (by $\Delta\kappa \lsim 0.02$
between $am_0=0.06$ and 0.02). Their position is quite independent of the
lattice size, as usual for sufficiently strong first order phase transitions.
These two observations allow us to estimate the positions of the chiral phase
transition in the $am_0=0$ limit (by linear extrapolation) and infinite volume
(by neglecting a volume dependence). The results are indicated by two circles
on the ET line in fig.~\ref{PD}.\footnote{The positions of the $1^{\rm st}$
  order transitions to the right and below the point T in fig.~\ref{PD} have
  been determined in a similar way.}

\begin{figure}%[t]
  \centerline{\epsfig{file=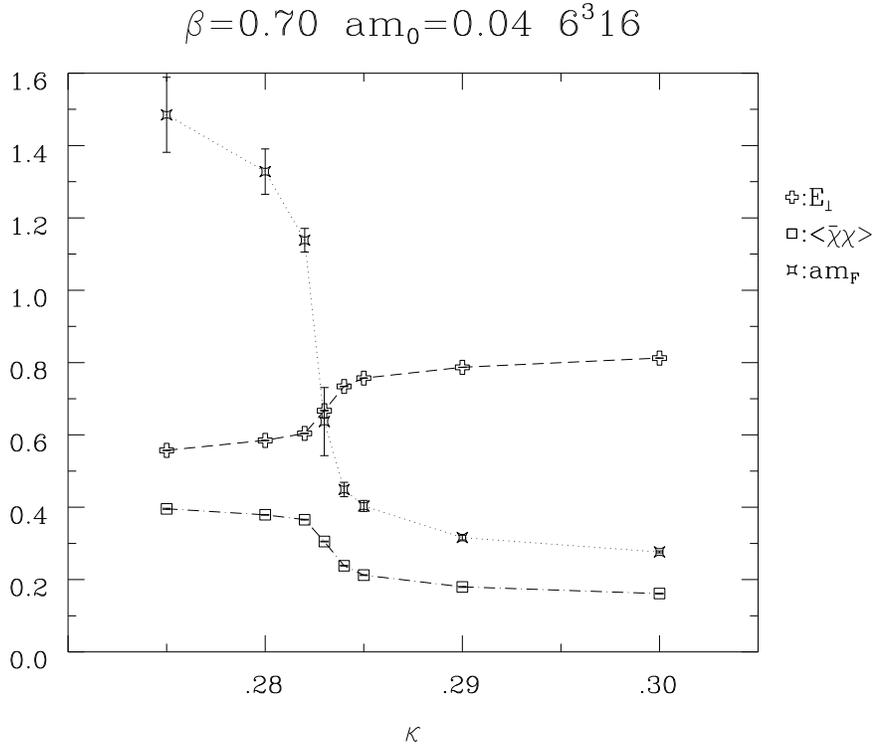,width=11cm,angle=90}}
  \vspace{-10mm}%
  \caption{%
    Energy $E_\perp$ (eq.~(\eq{E-PERP})), $\cbcex$ and $am_F$ as functions of
    $\kappa$ at $\beta=0.70$ for $am_0= 0.04$. The run at $\kappa=0.283$
    contains phase flips.}
  \label{fig:DISCO}
\end{figure}

To illustrate the coincidence of transitions in bosonic and fermionic
observables more precisely, we show in fig.~\ref{fig:DISCO} the observables
$E_\perp$ and $\cbcex$ and also $am_F$, all obtained in high statistics runs
(1600-9600 trajectories), plotted as functions of $\kp$ at $\bt = 0.70$ for
$am_0 = 0.04$. Again, the positions of the transitions agree within their
accuracy on a finite lattice. At $\kappa=0.283$ we have observed 2-3 phase
flips in $E_\perp$ and $\cbcex$, and the measured values of the observalbes
are thus to some extent accidental.

\begin{figure}%[t]
  \centerline{\epsfig{file=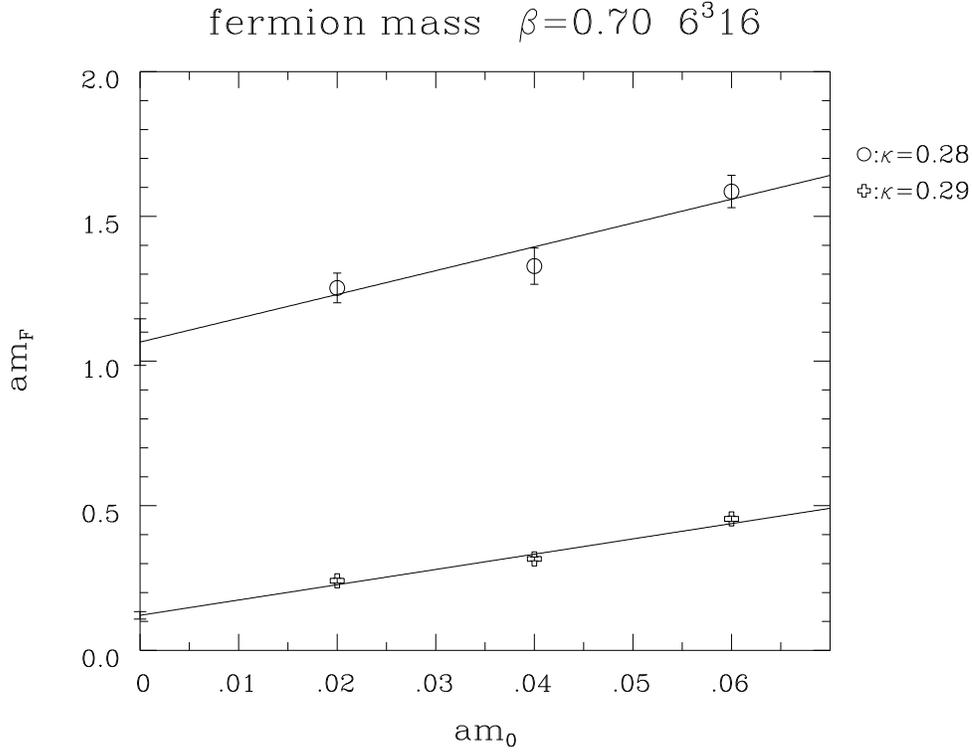,width=11cm,angle=90}}
  \vspace{-10mm}%
  \caption{%
    Fermion mass as a function of $am_0$ at $\beta=0.70$ for $\kappa=0.28$ and
    $\kappa=0.29$ on the $6^316$ lattice with a linear extrapolation to
    $am_0=0$.}
  \label{fig:mf_disc}
\end{figure}
In order to demonstrate that $am_F$ has a discontinuity also in the chiral
limit we have linearly extrapolated $am_F$, obtained in long runs, to $am_0=0$
very close below and above the phase transition, as shown in
fig.~\ref{fig:mf_disc}. The value of $am_F$ at $\beta=0.70$ on the $6^316$
lattice drops from $\simeq 1$ in the Nambu phase to nearly 0 in the Higgs
phase.

The agreement of the positions of discontinuities of bosonic and fermionic
observables suggests that the positions of the Higgs and chiral
phase transitions on the ET line coincide.  We conjecture that this is so, but
we have to admit that the data leave open a logical possibility that the
fermion observables do not jump exactly to zero as $\kp$ increases across the
Higgs phase transition.  The genuine chiral phase transition might take place
at some larger $\kp$, with small but finite values of fermionic observables in
between, hidden by finite lattice size, finite $am_0$, etc., effects.
Therefore the -- very crucial -- conjecture of the coincidence of both
transitions should be always checked in future studies with larger resources.

%44444444444444444444444444444444444444444444444444444444444444444444444
%%%%%%%%%%%%%%%%%%%%%%%%%%%%%%%%%%%%%%%%%%%%%%%%%%%%%%%%%%%%%%%%%%%%%%%%
%%%%%%%%%%%%%%%%%%%%%%%%%%%%%%%%%%%%%%%%%%%%%%%%%%%%%%%%%%%%%%%%%%%%%%%%
\section{NE line}

\begin{figure}%[t]
  \centerline{\epsfig{file=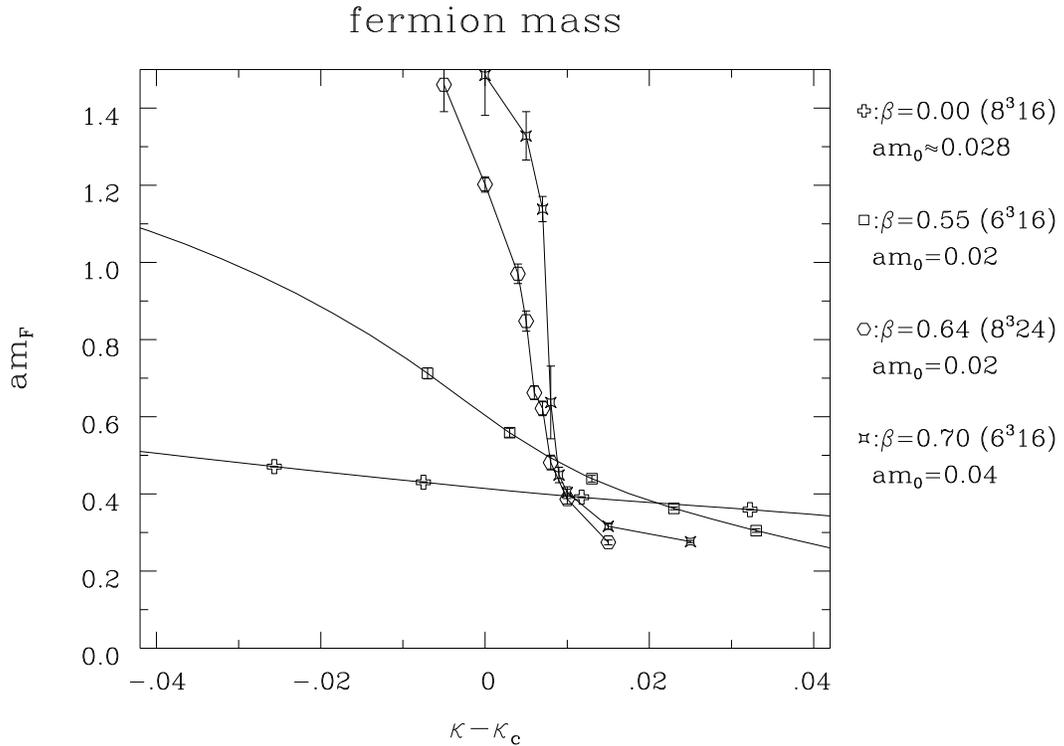,width=11cm,angle=90}}
  \vspace{-10mm}%
  \caption{%
    Comparison of the dependence of the fermion mass $am_F$ on $\kappa$ for
    various $\beta\geq 0$. $\kappa_c$ is the extrapolated critical point at
    $am_0=0$ in infinite volume (see table~\ref{PDtab}). The data for
    $\beta=0$ (crosses) are taken from ref.~\protect\cite{AlGo95}.}
  \label{fig:k_mf}
\end{figure}

The NE line is apparently of second order, as we have found no indications of
metastability in the range $0\leq \beta \leq 0.60$. This was to be expected
for small $\beta$ where the chiral transition is essentially that of the NJL
model~\cite{LeShr87b}.  As $\beta$ increases the chiral transition gets
substantially steeper, as demonstrated in fig.~\ref{fig:k_mf}.

At $\beta = 0$, we have checked that our data agree within good accuracy with
high precision data obtained recently~\cite{AlGo95} for the NJL model.  This
is not completely trivial, as we use a form of the Lee--Shrock transformation,
which is different from that applied in \cite{AlGo95} (no scalar field is used
there, and the model after the transformation is not gauge invariant).

Our aim was to find out how far the critical behavior in the vicinity of the
NE line is analogous to that at $\bt = 0$, i.e.\ to the NJL model, when $\beta$
is increased.  That this must be so for some finite interval in $\bt$ around
$\bt = 0$ follows from the strong coupling expansion \cite{LeShr87b}.  Our
strategy is to compare our data at various $\bt > 0$ with the data obtained in
ref.~\cite{AlGo95} at $\bt = 0$ using the same analytic Ans\"atze.

\begin{figure}%[t]
  \centerline{\epsfig{file=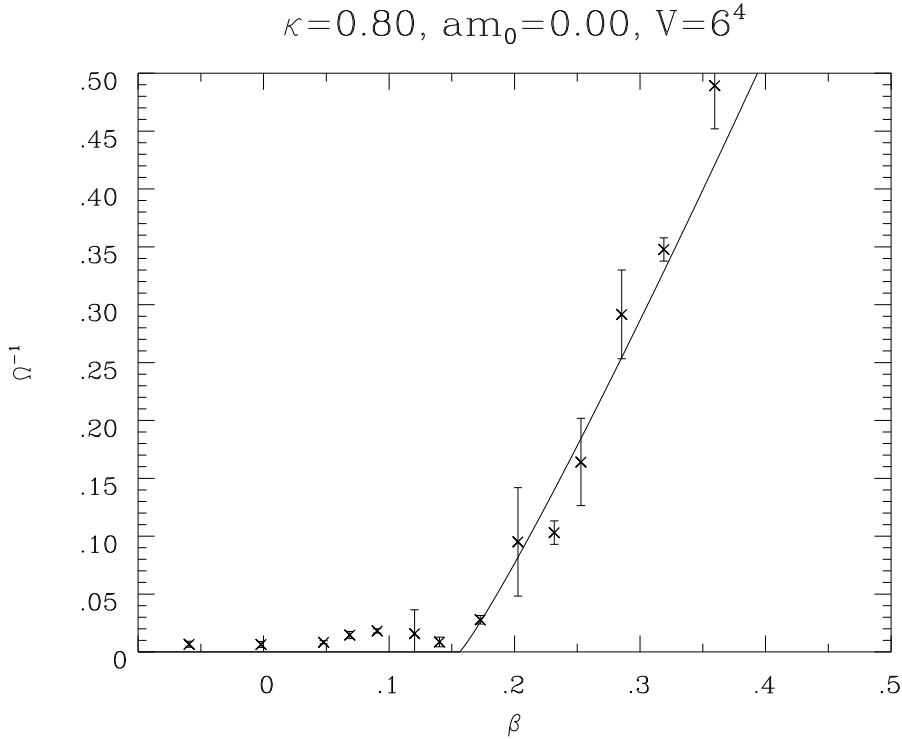,width=11cm,angle=90}}
  \vspace{-10mm}%
  \caption{%
    The observable $\Omega^{-1}$ defined in eq.~(\ref{eq:omega}). The curve is a
    power law fit described in the text.}
  \label{fig:omega}
\end{figure}

The position of the phase transition has been estimated on the $6^4$ lattice
using the microcanonical fermionic average method~\cite{Lu95}, allowing a
calculation at $am_0=0$, but without an extrapolation to the infinite volume.
The method is described in the appendix.  The observable $\Omega$,
eq.~(\eq{eq:omega}), obtained by this method at fixed $\kappa=0.80$, is shown
in fig.~\ref{fig:omega}.
The positions of the pseudocritical points $\beta_{pc}$ on $6^4$ lattice have
been determined by means of a power law fit of the form $\Omega^{-1}(\beta) =
A (\beta-\beta_{pc})^\gamma$ to the significantly nonzero values of
$\Omega^{-1}$. The obtained values of $\gamma$ are consistent with 1 within
the systematic error $\pm0.2$, estimated by fits to different sets of points.
The resulting $\beta_{pc}$ are indicated in fig.~\ref{PD} by crosses.

The position $\kappa_c(\beta)$ of the chiral phase transition at $am_0=0$ on
the infinite volume lattice has been determined by means of the hybrid Monte
Carlo algorithm at finite $am_0$ and on various lattice sizes. The
extrapolation to the infinite volume and $m_0=0$ has been performed by means
of a modified gap equation at $\beta=0$, 0.55 and at $\kappa=0.60$ (3 circles
most to the left in fig.~\ref{PD}) as explained below. The results of the
extrapolation are listed in table~\ref{PDtab}.

The modified gap equation has been used in ref.~\cite{AlGo95} for analytic
description of the data for the chiral condensate in the NJL model on
different lattices and for different $am_0$. It is defined as follows:
\begin{equation}
  \cbcex = \frac{1}{V} \sum_{k} \frac{J}{J^2+c_1^2\sum_\mu \sin^2 k_\mu}
  \lb{eq:modgap}
\end{equation}
with
\begin{equation}
  J = c_1^2 w(\kappa,\beta) \cbcex + c_2 am_0\,.
\end{equation}
This equation is a generalization of the usual gap equation by introducing
free parameters $w(\kappa,\beta)$, $c_1$ and $c_2$ independent of $am_0$ and
lattice size.  The $c_1$ and $c_2$ are expected to be nearly independent of
$\kappa$ for $\kappa\simeq \kappa_c$. In the original gap equation for the NJL
model their values are $w(\kappa,0)=8G(\kappa)$, $c_1=r(\kappa)$ and $c_2=1$
(for the definition of $G$ and $r$ see eq.~(\ref{rofkappa}) and (\ref{GNJL})).
For $\kappa\simeq\kappa_c$ the variation of $r(\kappa)$ is small at $\beta=0$.
To use the modified equation also for $\beta>0$ we neglect this presumably
weak dependence and keep $c_1$ constant. Up to this simplification and some
redefinitions of the parameters the equation~(\ref{eq:modgap}) is the same as
eq.~(3.4) of ref.~\cite{AlGo95}. We can thus compare our data analysis at
$\beta>0$ with that performed in \cite{AlGo95} at $\beta=0$.

We have used eq.~(\eq{eq:modgap}) as an implicit equation for $w$. For each
value of $\beta$ we choose $c_1$ and $c_2$, and then for each
$\kappa\simeq\kappa_c$, $am_0$ and lattice size we determine $w$ from the
value of $\cbcex$.  The choice of $c_1$ and $c_2$, independent of $\kappa$,
$am_0$ and lattice size, is made so that $w$ is at fixed $\beta$ independent
of $am_0$ and lattice size as much as possible. We minimize the expression
\begin{equation}
  \sum_i\left(w_i - \overline{w}(\kappa,c_1,c_2)\right)^2 \frac{1}{\sigma_i}
  \lb{eq:sigw}
\end{equation}
with respect to variations of $c_1$ and $c_2$. Here $w_i$ and $\sigma_i$ are,
respectively, values and errors of $w$ obtained by means of (\eq{eq:modgap})
from $\cbcex$ value and its statistical error at different data points $i$.
$\overline{w}$ is the average of $w_i$ for the same $\kappa$, $c_1$ and $c_2$.

\begin{figure}%[t]
  \vspace{-8mm}%
  (a)\hspace{15mm}%
  \parbox[c]{13cm}{\epsfig{file=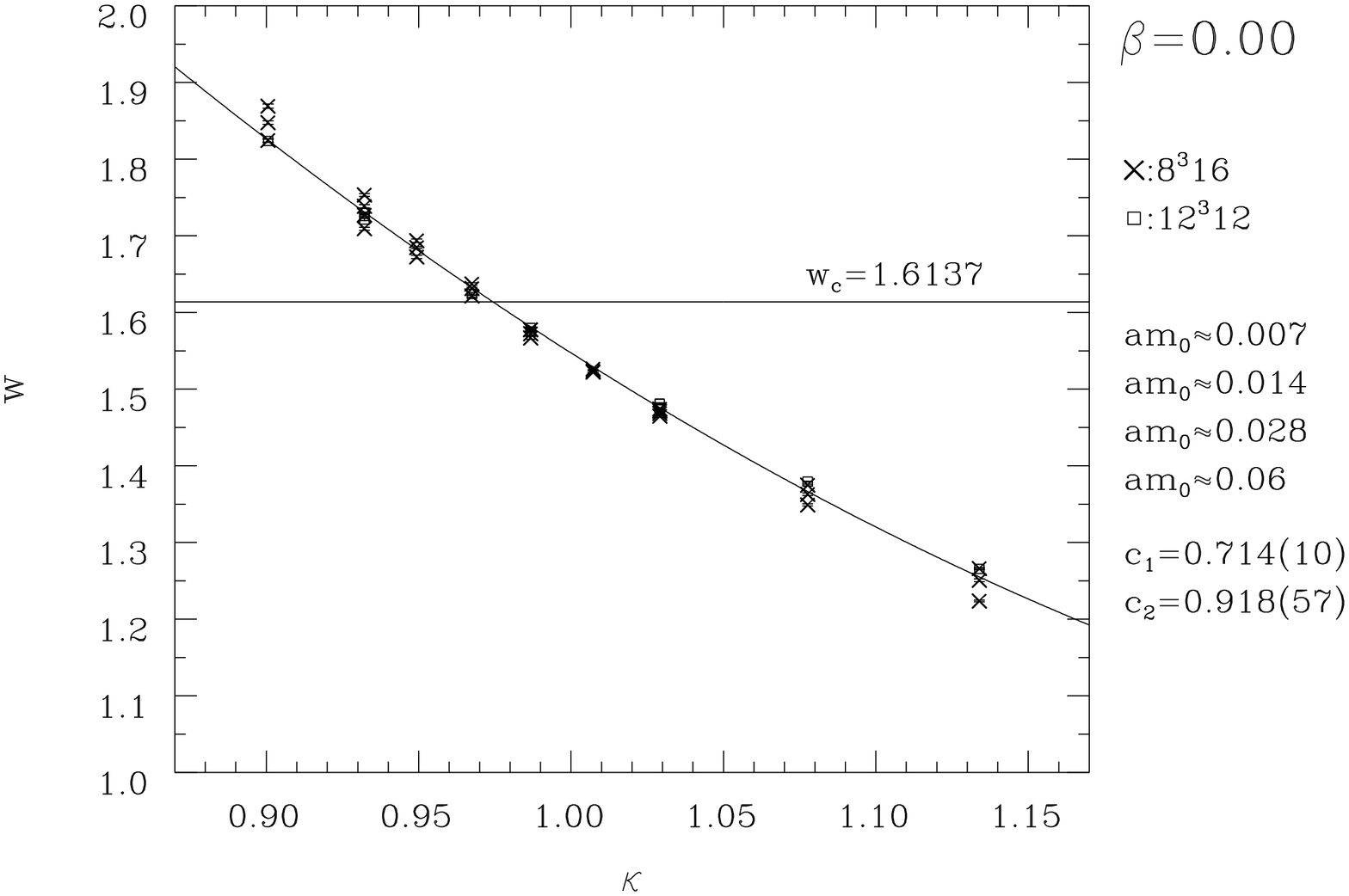,width=8.8cm,angle=90}}\\[-13.5mm]
  (b)\hspace{15mm}%
  \parbox[c]{13cm}{\epsfig{file=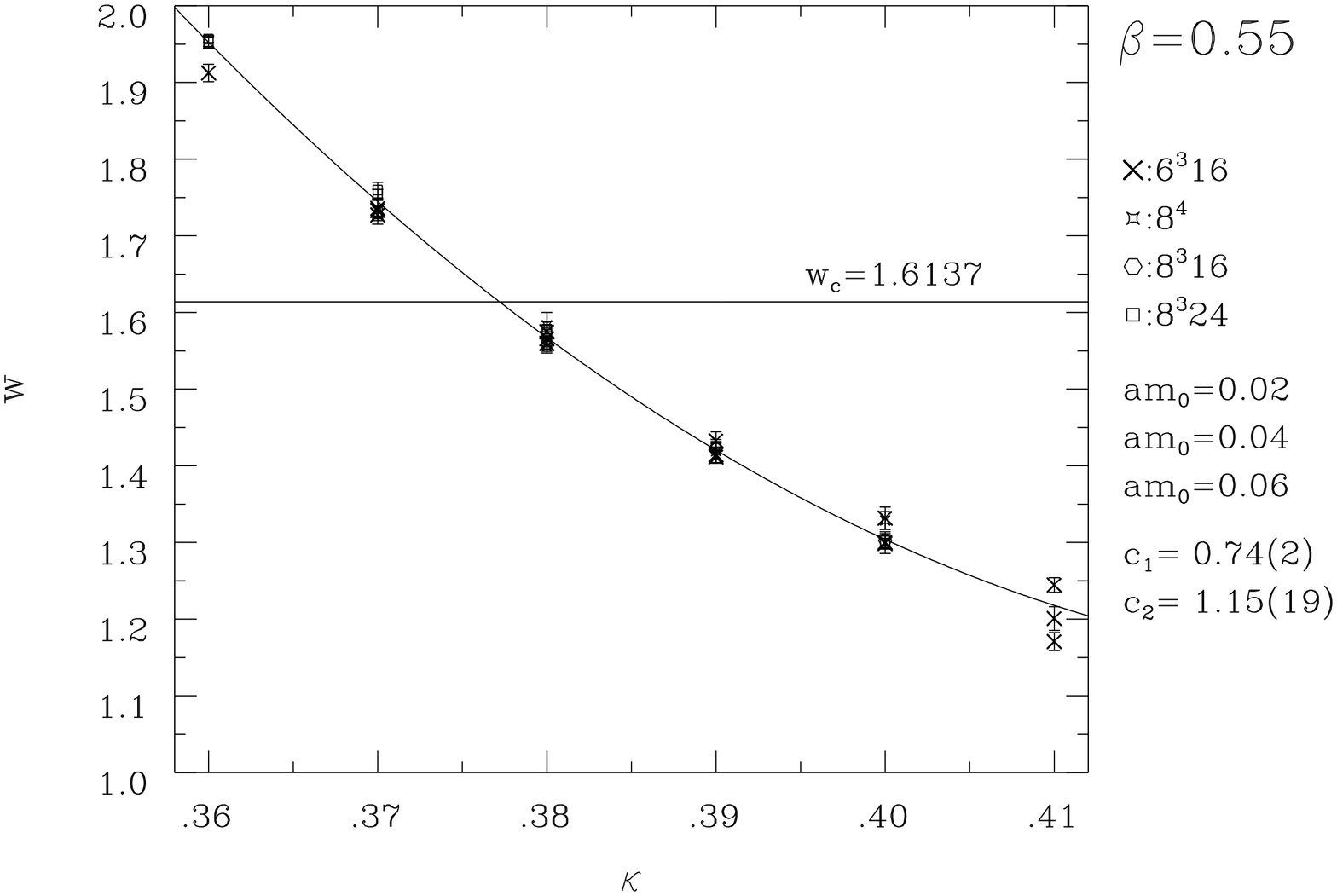,width=8.8cm,angle=90}}\\[-13.5mm]
  (c)\hspace{15mm}%
  \parbox[c]{13cm}{\epsfig{file=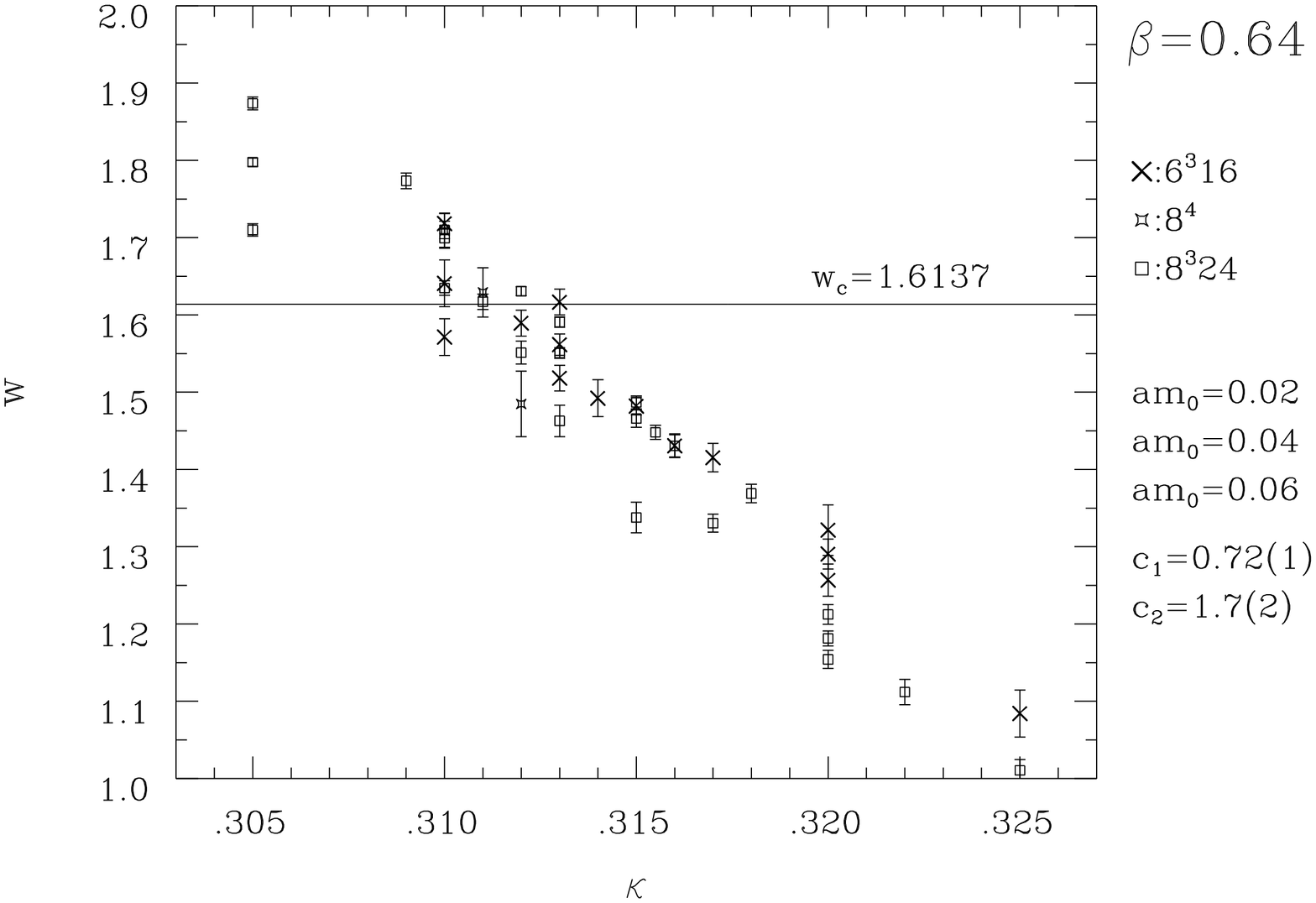,width=8.8cm,angle=90}}
  \vspace{-9mm}%
  \caption{\label{fig:ige}%
    The parameter $w$ in the modified gap equation as a function of $\kappa$
    for (a) $\beta=0$ (NJL model), (b) $\beta=0.55$ and (c) $\beta=0.64$.  For
    $\beta=0$ and $\beta=0.55$ $w$ is independent of lattice sizes and bare
    mass and the modified gap equation describes the data well. For
    $\beta=0.64$, which is close to the point E, this is no longer so.  The
    curves in (a) and (b) are quadratic function fits to the $w$-values. The
    data in figure (a) are taken from ref.~\cite{AlGo95}.}
\end{figure}

The possibility of choosing $c_1$ and $c_2$ in such a way that $w$ depends only on
$\kappa$ means that the modified gap equation is applicable at this $\beta$.
The results for $w(\kappa)$ are shown in fig.~\ref{fig:ige}a,b.
If the modified gap equation works, it can be used for an extrapolation to
chiral limit and infinite volume, and the critical value $w_c$ of $w$, below
which $\cbcex$ vanishes (see fig.~\ref{fig:ige}), is obtained. The critical
point $\kappa_c(\beta)$ at the given $\beta$ is then determined from
$w(\kappa_c)=w_c$.

\begin{figure}%[t]
  \centerline{\epsfig{file=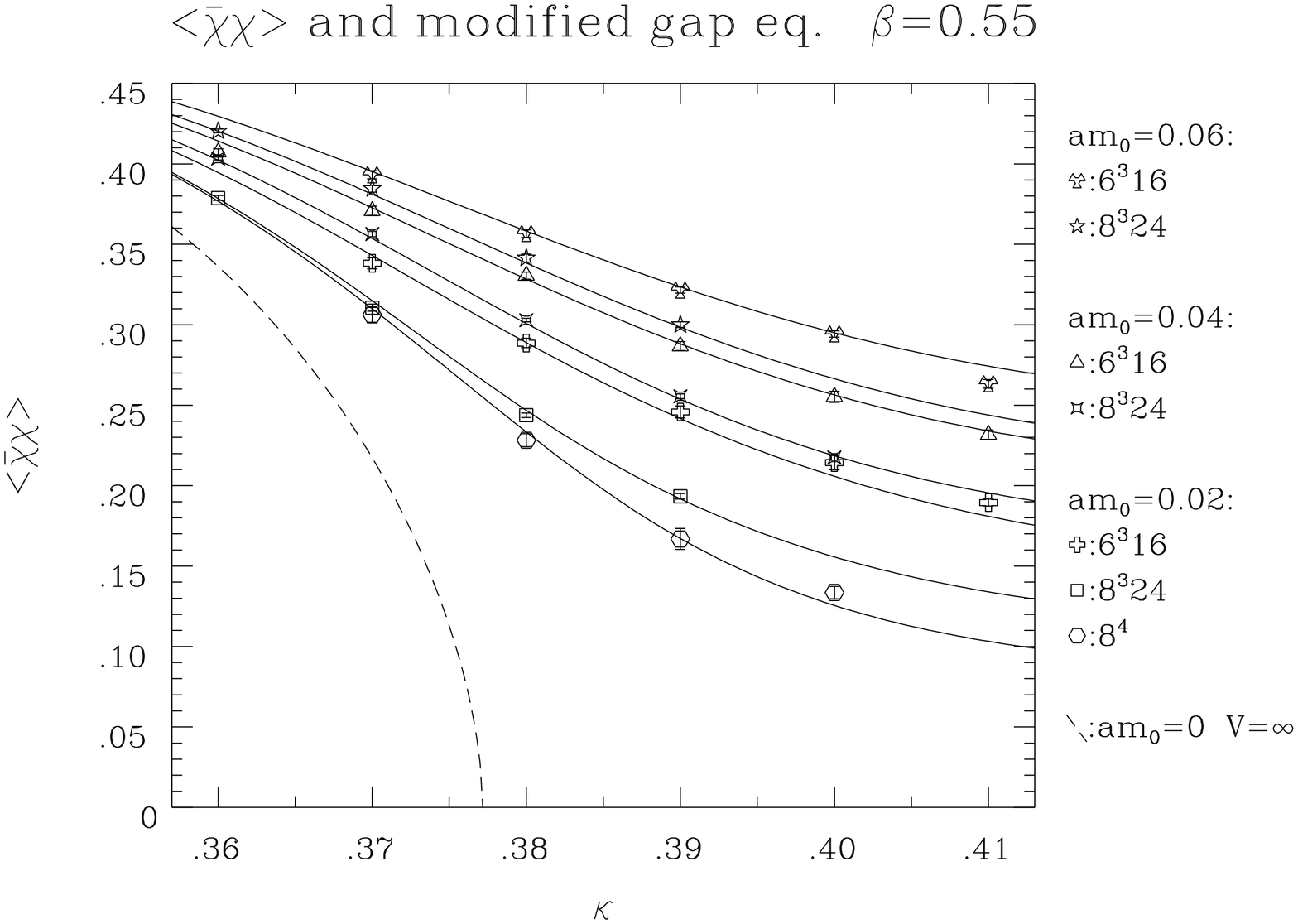,width=11cm,angle=90}}
  \vspace{-10mm}%
  \caption{%
    Comparison of the modified gap equation and the data at $\beta=0.55$. The
    parameter $w$ depends on $\kappa$ as described by the line in
    fig.~\protect\ref{fig:ige}b, whereas the parameters $c_1$ and $c_2$ are
    fixed at values $c_1=0.74$ and $c_2=1.1$. The extrapolation to infinite
    volume and $am_0=0$ is represented by the dashed line.}
  \label{fig:modgap}
\end{figure}

As seen from fig.~\ref{fig:ige}a and \ref{fig:ige}b, the modified gap equation
describes in the above sense the data very well at $\beta=0$ and $\beta=0.55$
for various $am_0$ and different lattice sizes.  The agreement at $\beta=0.55$
(fig.~\ref{fig:modgap}) is nearly as good as at $\beta=0$, cf.~ref.\ 
\cite{AlGo95}. The constants $c_1$ and $c_2$ have nearly the same values for
these two $\beta$ values. Therefore this method can be used also for fixed
$\kappa$ and varying $\beta\simeq\beta_c$.  We have done this at
$\kappa=0.60$.
The lowest curve in fig.~\ref{fig:modgap} describes the extrapolation to
infinite volume and $am_0=0$. In this way the position $\kappa_c(\beta)$
of the chiral transition has been obtained for 3 points on the NE line.

In this way one can obtain also pseudocritical points on lattices of finite
size, again by determining values of $w$ for which $\cbcex$ vanishes. The
corresponding $\beta_{pc}$ on the $6^4$ lattice are consistent with those
obtained by the microcanonical fermion average method.

\begin{figure}%[t]
  (a)\parbox[c]{15.8cm}{\epsfig{file=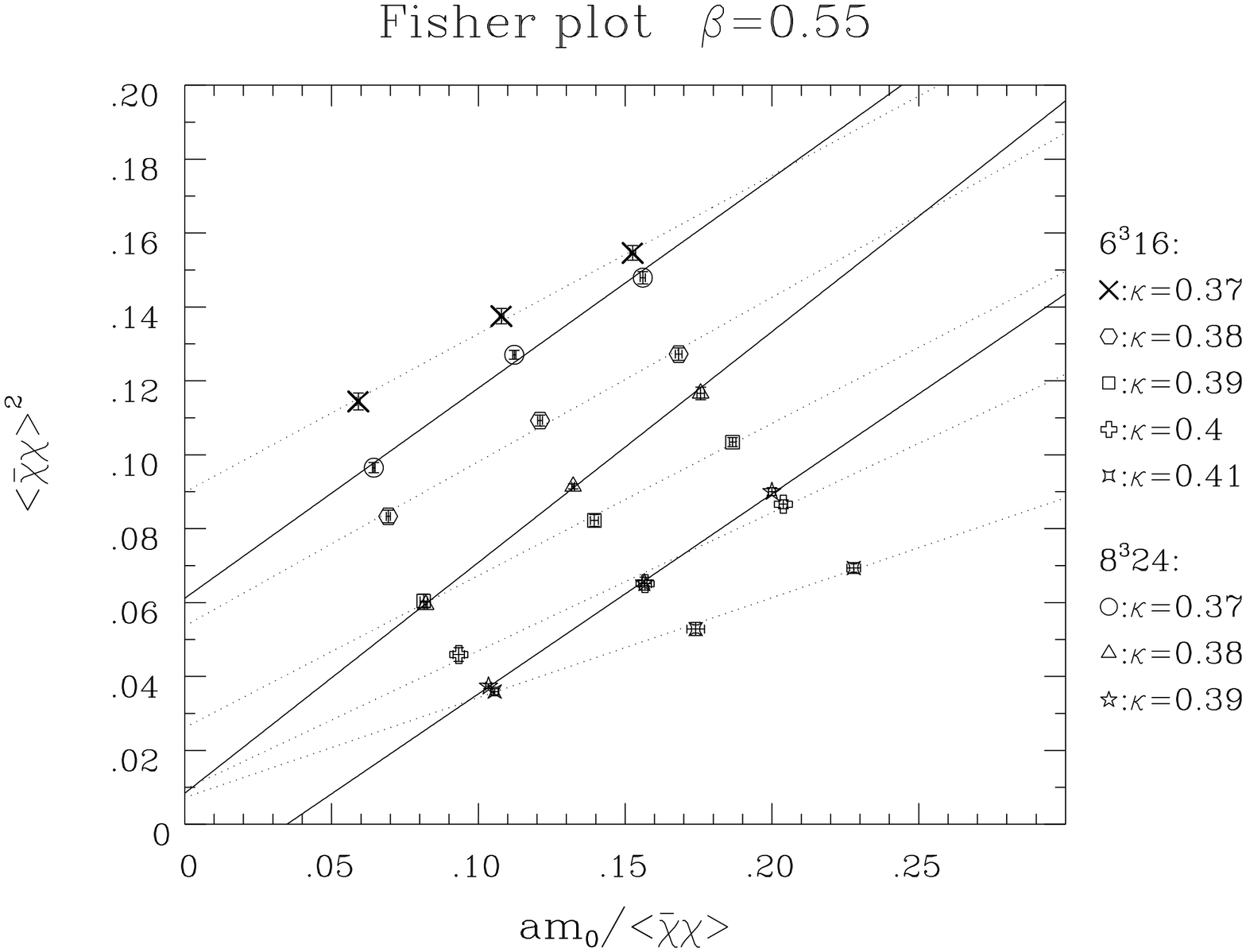,width=11cm,angle=90}}\\[-5mm]
  (b)\parbox[c]{15.8cm}{\epsfig{file=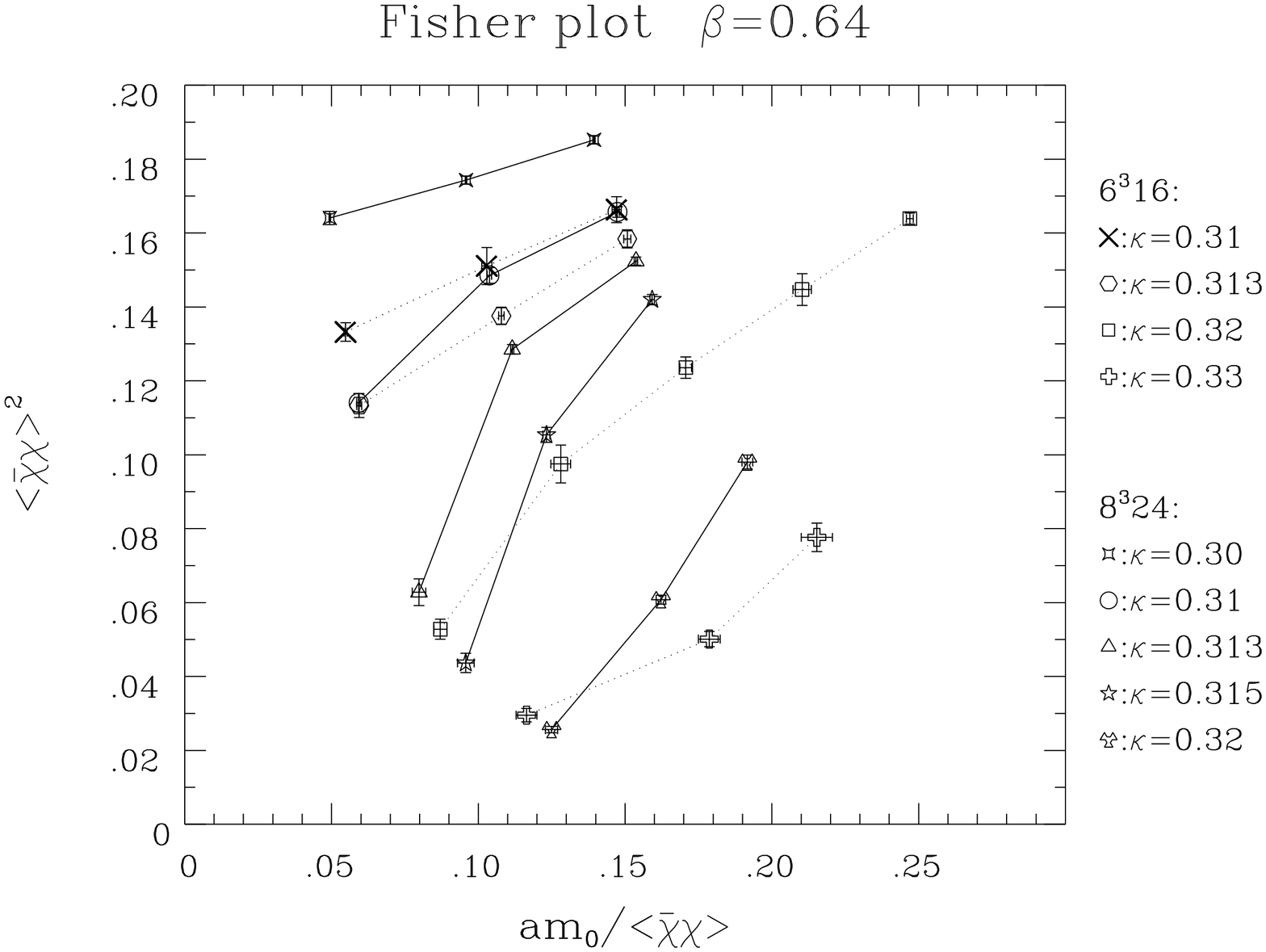,width=11cm,angle=90}}%
  \vspace{-10mm}%
  \caption{%
    $\cbcex^2$ as a function of $am_0/\cbcex$ for fixed coupling parameters
    and lattice size (Fisher plots) at (a) $\beta=0.55$ and (b)
    $\beta=0.64\simeq\beta_{\rm E}$. The straight lines in (a) indicate
    according to eq.~(\ref{fisher}) a mean field behavior.}
  \label{fig:fisher}
\end{figure}

Another way to investigate the scaling properties of the NJL model in
\cite{AlGo95} was to plot $\cbcex^2$ as a function of $am_0/\cbcex$
(Fisher plot).  This is motivated by the general scaling relation
\begin{equation}
  \label{fisher}
  am_0 = b_1 (\kappa-\kappa_c) \cbcex^{\left(\delta-\frac{1}{\beta}\right)}
  + b_2 \cbcex^\delta\;,
\end{equation}
where the mean field values of the exponents, $\delta=3$ and
$\beta=\frac{1}{2}$, are assumed. Provided these values are correct, the data
for a fixed $\kappa$ on a given lattice should form straight lines.
As seen in fig.~\ref{fig:fisher}a, for $\beta=0.55$ the mean field exponents
describe the data quite well. 
We again find a large similarity between our data at $\beta=0.55$ and the data
at $\beta=0$ in ref.~\cite{AlGo95}.  This includes also the observed finite size
effect.

%55555555555555555555555555555555555555555555555555555555555555555555555
%%%%%%%%%%%%%%%%%%%%%%%%%%%%%%%%%%%%%%%%%%%%%%%%%%%%%%%%%%%%%%%%%%%%%%%%
%%%%%%%%%%%%%%%%%%%%%%%%%%%%%%%%%%%%%%%%%%%%%%%%%%%%%%%%%%%%%%%%%%%%%%%%
%%%%%%%%%%%%%%%%%%%%%%%%%%%%%%%%%%%%%%%%%%%%%%%%%%%%%%%%%%%%%%%%%%%%%%%%
\section{Vicinity of the point E}
It is notoriously difficult to localize a tricitical point. Our best current
estimate is $\beta_{\rm E} = 0.64^{+0.03}_{-0.04}$ and $\kappa_{\rm E} =
0.305^{+0.03}_{-0.015}$ (point E in fig.~\ref{PD}). It is based on the
extrapolation in $\beta$ of latent heat to zero and on the localization of
minimum of scalar mass $am_S$ as described below. Finite size effects are only
roughly taken into account.

We have tried to apply at $\beta=0.64$ the same method of analysis to the
scaling behavior as at $\beta=0$ and $\beta=0.55$. It turns out that both
above methods do not work anymore. The modified gap equation does not describe
the data if $w$ should depend only on $\kappa$ and not on $am_0$ and lattice
size.  This is seen in figure~\ref{fig:ige}c. Here we show the values of $w$
obtained for such a set of parameters $c_1$ and $c_2$, for which the
dependence of $w$ on $am_0$ and lattice size is minimal.  Also the test for
mean field values of exponents $\beta$ and $\delta$ by means of the Fisher
plot fails. In fig.~\ref{fig:fisher}b, the lines connecting data at the same
$\kappa$ and lattice size differ significantly from straight lines.

We have found another difference between the behavior of the model close to
the chiral phase transition around the point E and at smaller $\beta$. The
scalar boson mass $am_S$ gets small in the vicinity of E, whereas it remains
of the order of cutoff or larger at the chiral transition for smaller $\beta$.
\begin{figure}%[t]
  (a)\parbox[c]{15.8cm}{%
    \epsfig{file=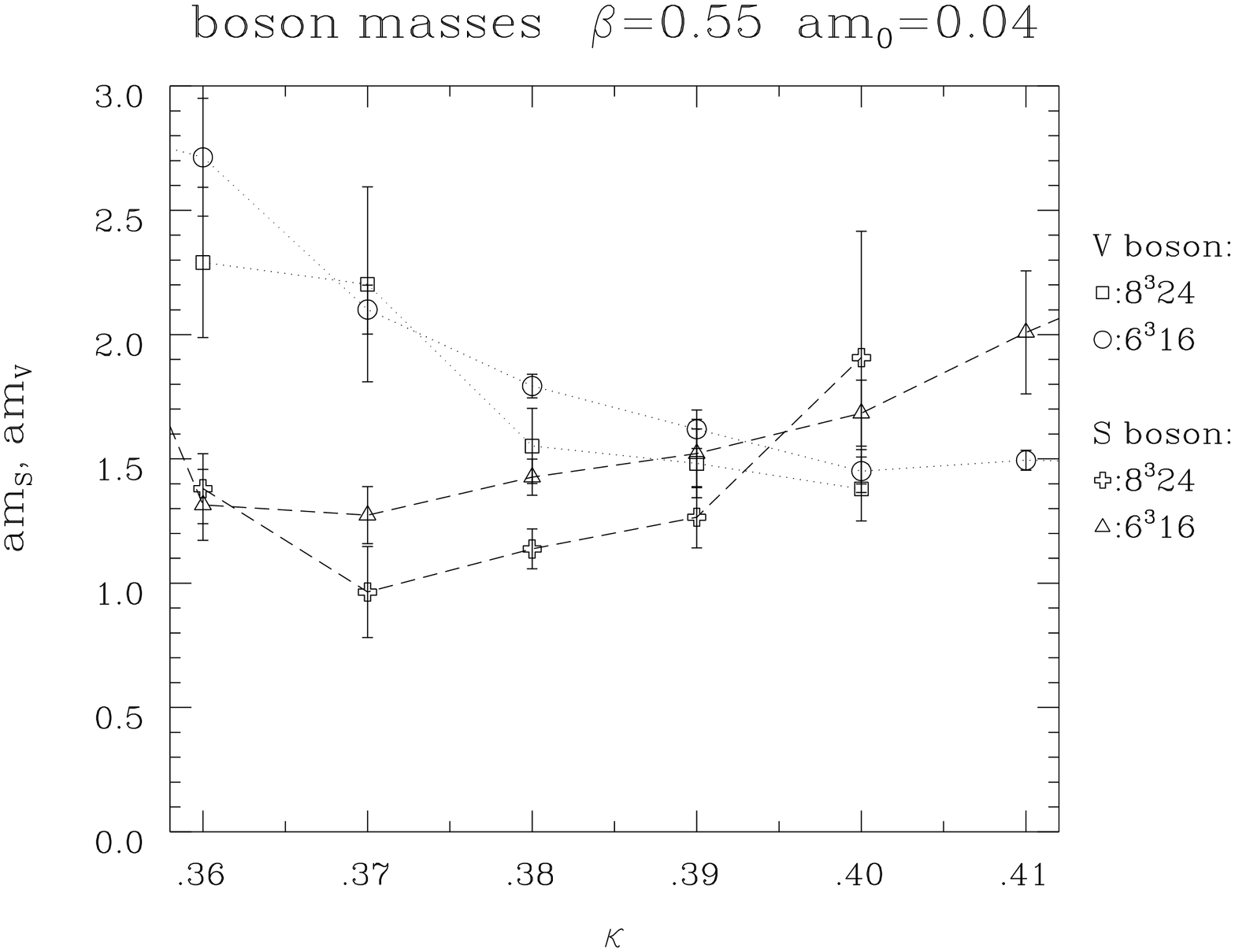,width=11cm,angle=90}}\\[-5mm]
  (b)\parbox[c]{15.8cm}{%
    \epsfig{file=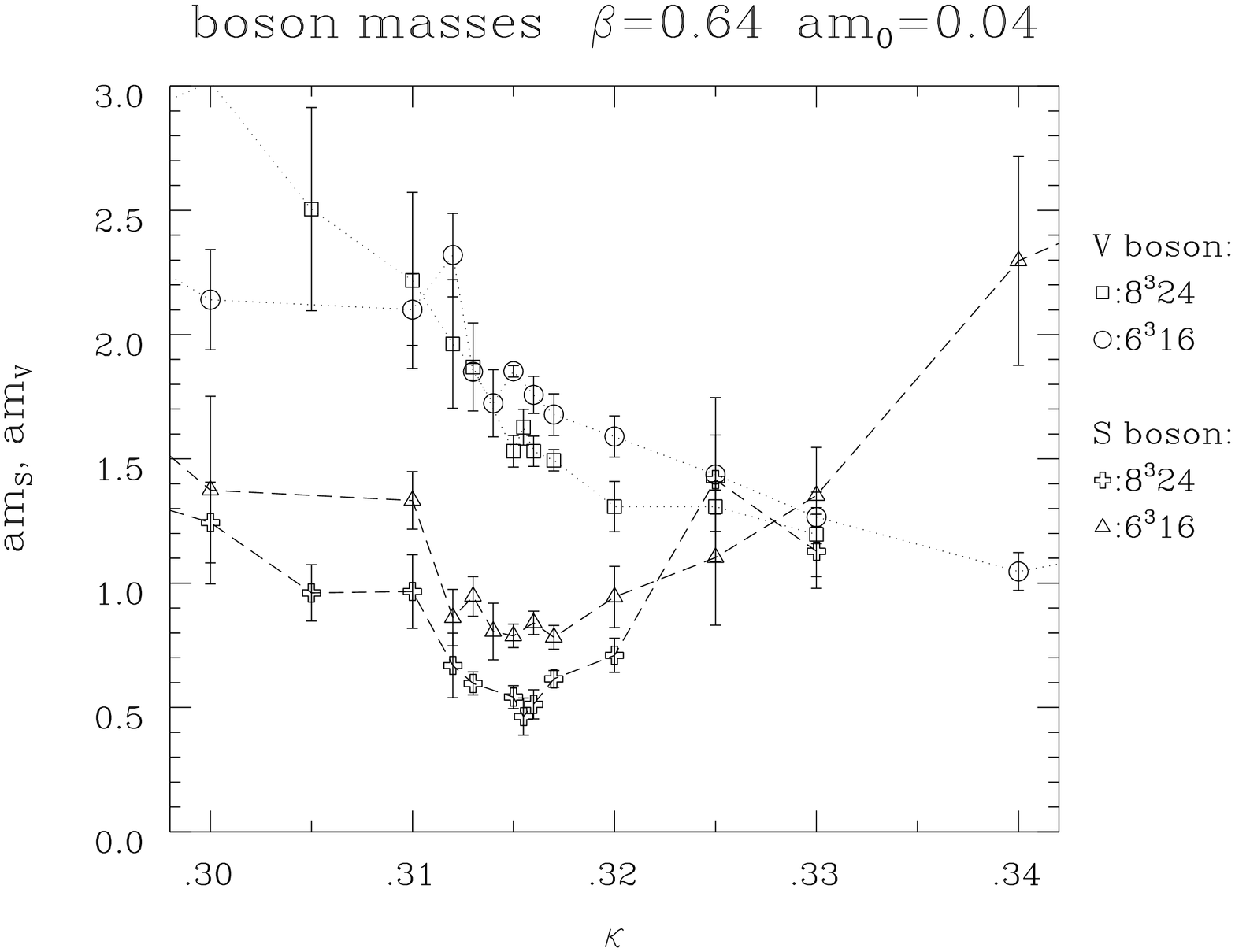,width=11cm,angle=90}}%
  \vspace{-10mm}%
  \caption{%
    Vector and scalar boson masses, $am_V$ and $am_S$, as functions of
    $\kappa$ at (a) $\beta=0.55$ and (b) $\beta=0.64\simeq\beta_{\rm E}$. To
    estimate the finite size effects two different lattice sizes ($6^316$ and
    $8^324$) are shown.}
  \label{fig:boson}
\end{figure}
In fig.~\ref{fig:boson}, both boson masses $am_S$ and $am_V$ are shown at
$\beta=0.55$ and $\beta=0.64$ for $am_0=0.04$. A dip in $am_S$ at $\beta=0.64$
around $\kappa=0.315$ is clearly seen. It gets deeper when the lattice size is
increased.  Similar dips have also been found for other values of $am_0$ and
lattice sizes for various $\beta\simeq \beta_{\rm E}$, whereas at $\beta=0.55$
or smaller $\beta$ the mass $am_S$ does not decrease below 1 neither on small
lattices nor on large ones.  This difference, obvious in the quenched case, is
thus present also with dynamical fermions. If a new correlation length
diverges at some critical point and not at other ones, the most natural
explanation is that the scaling behavior is different.

These observations indicate that the scaling behavior in the vicinity of the
point E is different from that at smaller $\beta$. This tentative conclusion
is supported by the experience with tricritical points in various statistical
mechanics models~\cite{LaSa84}. The tricritical exponents have usually quite
different values from those describing the universality class of the adjacent
critical manifolds. We also note that the chiral and magnetic phase
transitions, occurring in the \chupiv\ model at small and large $\beta$,
respectively, meet at the point E.  Thus phenomena quite different from those
known in both limit cases might occur here.

%66666666666666666666666666666666666666666666666666666666666666666666666
%%%%%%%%%%%%%%%%%%%%%%%%%%%%%%%%%%%%%%%%%%%%%%%%%%%%%%%%%%%%%%%%%%%%%%%%
%%%%%%%%%%%%%%%%%%%%%%%%%%%%%%%%%%%%%%%%%%%%%%%%%%%%%%%%%%%%%%%%%%%%%%%%
%%%%%%%%%%%%%%%%%%%%%%%%%%%%%%%%%%%%%%%%%%%%%%%%%%%%%%%%%%%%%%%%%%%%%%%%

\section{Summary and outlook}

From the study of the behavior of several observables in the vicinity of the
point E, as well as from some more general considerations about tricritical
points, we conclude that the tricritical point E in the phase diagramm of the
\chupiv\ model is an interesting candidate for an investigation of the
continuum limit of this model. The significant differences from the NJL-type
model suggest that a continuum theory with massive fermions might arise which
does not belong to the universality class of the Yukawa models in the sense of
ref.~\cite{HaHa91}.

\begin{figure}%[t]
  \centerline{\epsfig{file=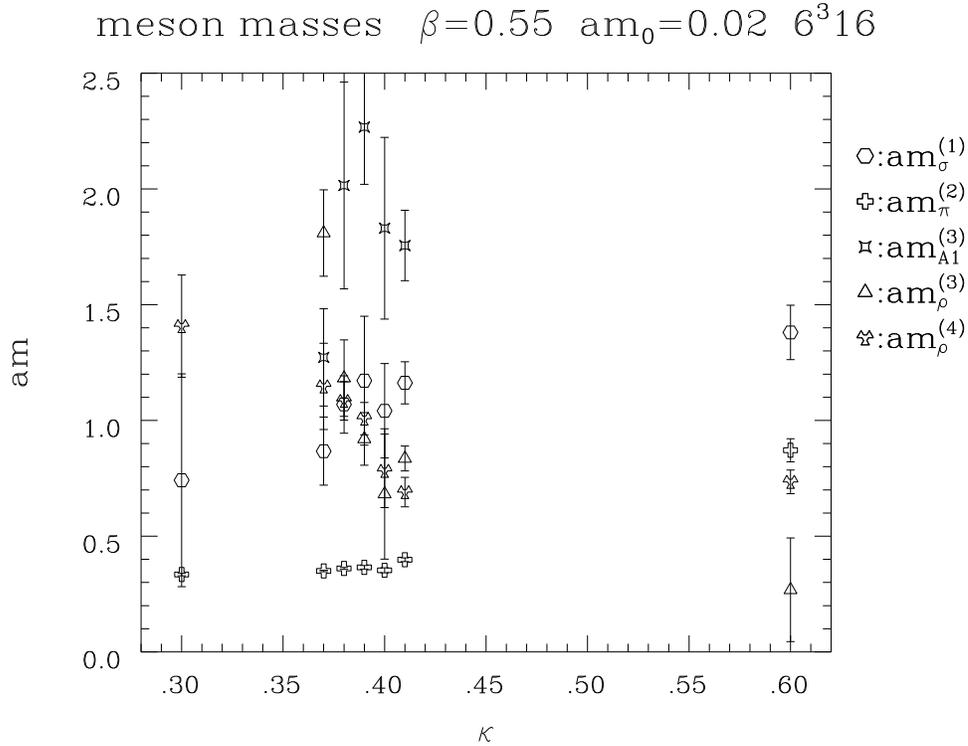,width=11cm,angle=90}}%
  \vspace{-10mm}%
  \caption{%
    Meson states observable in the vicinity of the chiral phase transition.
    The states are denoted according to ref.~\protect\cite{GoSm84}.}
  \label{fig:meson}
\end{figure}

To test this conjecture a laborious investigation of the scaling behavior of the
model around the tricritical point E will be necessary, taking more
observables into account. A rich spectrum is expected in the Nambu phase of
the model, qualitatively similar to that one would guess in the case of QCD
with one fermionic and one scalar quark (this analogy has been discussed in
ref.~\cite{FrJe95a}). Some glimpse at the states not too difficult to find is
provided by figs.~\ref{fig:boson} and \ref{fig:meson}.
In fig.~\ref{fig:meson} we show some meson masses. It has been checked
in~\cite{LuFr95} that for the Goldstone boson mass $am_\pi$ the relationship
$am_\pi \propto \sqrt{am_0}$, expected in the Nambu phase, holds. In this
figure one can identify also states which are analogies of the $\rho$,
$\sigma$ and $A_1$ mesons in QCD.  Such states might accompany the massive
fermion $F$ and the scalar $S$ also in the continuum limit of the \chupiv\ 
model constructed at the tricritical point E. If so, the spectrum would be
rather different from that of the Yukawa models, supporting the conjecture of
a new universality class.

\subsection*{Acknowledgements}
We thank V. Azcoiti, M. G\"ockeler, R. Horsley, M. Lindner, G. Schierholz and
R.E. Shrock for discussions. We are grateful to the authors of
ref.~\cite{AlGo95} for many explanations and H.A. Kastrup for continuous
support.  Most computations have been performed on Cray YMP/864 at HLRZ
J\"ulich ($6^316$ lattice) and VPP 500/4 Landesrechner in Aachen ($8^324$
lattice).

%%%%%%%%%%%%%%%%%%%%%%%%%%%%%%%%%%%%%%%%%%%%%%%%%%%%%%%%%%%%%%%%%%%%%%%%
%%%%%%%%%%%%%%%%%%%%%%%%%%%%%%%%%%%%%%%%%%%%%%%%%%%%%%%%%%%%%%%%%%%%%%%%

\appendix
\section{Microcanonical fermionic average method extended to field theories
  with scalars}

To perform the calculation of $\Omega^{-1}$, eq.~(\ref{eq:omega}), in the
chiral limit the microcanonical fermionic average method developed for the
gauge-fermion systems~\cite{AzDi90,AzLa93} has been extended to the
fermion-gauge-scalar models~\cite{Lu95}.  Here we just outline the strategy.

The essential idea of the algorithm is the computation of the full effective
action
\begin{equation}
  S_{\rm eff}(E,m_0,N_f,\beta,\kappa)=- \ln M(E,\kappa) 
  - 6V \beta E + S_{\rm eff}^{F} (E,m_0,N_f,\kappa) 
  \lb{FULL}
\end{equation}
as a function of the pure gauge energy $E$ and other bare parameters like
$m_0$, $N_f$, $\beta$ and $\kappa$, where
\begin{equation}
  M(E,\kappa)= \int [dU] [d \phi][d \phi^{\dagger}] \delta [E_{\rm P}(U)-E]
  \exp[8V\kappa E_{\rm L}(U,\phi)]
  \lb{DENSITY}
\end{equation}
is the density of states and
\begin{eqnarray}
  S_{\rm eff}^{F}(E,m_0,N_f,\kappa)] & = & -\ln \left\langle \left[ \det
      \Delta(U) \right]^\frac{N_f}{4} \right\rangle_{E} \nonumber\\
  & = & - \ln {\int [dU] [d \phi][d \phi^{\dagger}] \delta [E_{\rm P}(U)-E] 
    \left[ \det \Delta(U) \right]^\frac{N_f}{4} \exp [8V \kappa E_{\rm L}(U)]
    \over M(E,\kappa)}\hspace{1cm}
  \lb{SF}
\end{eqnarray}
is the effective fermionic action.  $M(E,\kappa)$ in (\ref{DENSITY}) can be
easily computed by quenched simulation.  The main effort to be paid is to
calculate $S_{\rm eff}^{F}(E,m_0,N_f,\kappa)$ through (\ref{SF}) by means of
microcanonical and Monte Carlo simulations \cite{Lu95}.  For $N_f=4$,
$\langle\det \Delta(U)\rangle_{E}$ is the fermionic determinant averaged over
the configurations with the probability distribution

\begin{eqnarray}
  {\delta [E_{\rm P}(U)-E] \exp[8V\kappa E_{\rm L}(U,\phi)] \over M(E,\kappa)}.
  \lb{PDF}
\end{eqnarray}
From (\ref{SF}), one sees that the effective fermionic action $S_{\rm eff}^F$
does not depend on $\beta$ and its dependence on $m_0$ and $N_f$ can be easily
obtained, once the eigenvalues of the massless fermionic matrix for the
decorrelated configurations at fixed $E$ have been calculated by the Lanczos
algorithm.  Once $S_{\rm eff}(E,m_0,N_f,\beta,\kappa)$ is known, the
thermodynamical quantities such as $\langle E_{\rm P}\rangle$, $\langle E_{\rm
  L}\rangle$, $\langle \bar\psi \psi\rangle$ and $\Omega^{-1}$ can be obtained
by deriving the partition function
\begin{eqnarray}
Z = \int dE \exp\left[-S_{\rm eff}(E,m_0,N_f,\beta,\kappa)\right].
\lb{FUN}
\end{eqnarray}
or by saddle point analysis \cite{Lu95}.

In \cite{Lu95}, on $6^4$ lattice and at finite $am_0$, comparisons with the
results from the Hybrid Monte Carlo (HMC) method for a fermion-gauge-scalar
model were made, and the HMC data were well reproduced.  The advantage of the
algorithm, in addition to its lower computational costs in searching the
parameter space $(am_0,\beta)$ compared to the conventional algorithms,
is the accessibility of the chiral limit. Therefore the algorithm is very
suitable for the phase structure analysis.

While having demonstrated the advantages of the algorithm on the $6^4$
lattice, we would like to mention some open questions.  For large lattice
volumes, it is time and memory consuming to diagonalize the fermionic matrix.
For example, on the $8^4$ lattice, the amount of data required grows so fast
that the algorithm has not been well tested.  Although the Lanczos algorithm
is parallelizable \cite{AzLa93}, it is not easily vectorizable.  It is also
not easy to evaluate observables other than the thermodynamical quantities.
These problems are under further investigation.
 
%\clearpage

%%%%%%%%%%%%%%%%%%%%%%%%%%%%%%%%%%%%%%%%%%%%%%%%%%%%%%%%%%%%%%%%%%%%%%%%%%%%%%%%%%%%
%%%%%%%%%%%%%%%%%%%%%%%%%%%%%%%%%%%%%%%%%%%%%%%%%%%%%%%%%%%%%%%%%%%%%%%%%%%%%%%%%%%%

% \bibliography{jourabbr,net,our-papers,gauge,yukawa,referen}

% end_of_bibliography

\end{document}